\documentclass[%
 reprint,
 floatfix,
 amsmath,amssymb,
 aps,
 prd,
]{revtex4-2}

\usepackage{graphicx}
\usepackage{dcolumn}
\usepackage{bm}
\usepackage{hyperref}
\usepackage{subcaption}
\usepackage[english]{babel}
\usepackage{float}
\usepackage[section]{placeins}
\usepackage{algorithm}
\usepackage{algpseudocode}

\graphicspath{{./}}     

\begin{document}

\preprint{APS/PRD-QED}

\title{Model-Agnostic Population Inference for Gravitational-Wave Astronomy: From LVK to LISA}

\author{Yi-kun Li}
\affiliation{State Key Laboratory of Radio Astronomy and Technology, Xinjiang Astronomical Observatory, CAS, 150 Science 1-Street, Urumqi 830011, China}
\affiliation{School of Astronomy and Space Science, University of Chinese Academy of Sciences, No.19A Yuquan Road, Beijing, 100049, China}

\author{Cheng Cheng}
\affiliation{State Key Laboratory of Radio Astronomy and Technology, Xinjiang Astronomical Observatory, CAS, 150 Science 1-Street, Urumqi 830011, China}
\affiliation{School of Astronomy and Space Science, University of Chinese Academy of Sciences, No.19A Yuquan Road, Beijing, 100049, China}

\author{Yun Fang}
\email[Corresponding author: ]{fangyun@nbu.edu.cn, cuilang@xao.ac.cn}
\affiliation{Institute of Fundamental Physics and Quantum Technology, Ningbo University, Ningbo 315211, China}
\affiliation{School of Physical Science and Technology, Ningbo University, Ningbo 315211, China}

\author{Lang Cui}
\email[Corresponding author: ]{fangyun@nbu.edu.cn, cuilang@xao.ac.cn}
\affiliation{State Key Laboratory of Radio Astronomy and Technology, Xinjiang Astronomical Observatory, CAS, 150 Science 1-Street, Urumqi 830011, China}
\affiliation{School of Astronomy and Space Science, University of Chinese Academy of Sciences, No.19A Yuquan Road, Beijing, 100049, China}

\date{\today}

\begin{abstract}
Characterizing compact-binary populations requires inference methods that can separate intrinsic structure from measurement uncertainty and detector selection effects without imposing an overly restrictive parametric form. We present a model-agnostic hierarchical population-inference framework based on a correlated compound-mixture population density and a normalizing-flow variational guide. The population model combines flexible bounded marginals, finite mixtures, and Gaussian-copula dependence, while the variational guide provides reusable posterior approximations for catalog-level hyperparameters. We validate the method in two detector regimes. For simulated Laser Interferometer Space Antenna (LISA) supermassive black hole binary catalogs, a truth-blind multistart repeated-validation study over 20 catalogs and 100 independent initializations supports recovery of both population shape and expected-count normalization in controlled simulations. For LIGO--Virgo--KAGRA (LVK) GWTC-3 binary black holes, we apply the same hierarchical selection-correction framework to GWOSC posterior samples and official O1/O2/O3 injection assets, and compare the inferred population shape with official GWTC-3 reference curves. Together, these applications show how flexible population densities can be combined with standard selection correction across distinct gravitational-wave detector regimes.
\end{abstract}

\maketitle

\section{Introduction}
\label{sec:introduction}

Gravitational-wave (GW) astronomy has become a catalog-based observational science. Since the first binary black hole detection \cite{Abbott2016}, the LIGO--Virgo--KAGRA (LVK) network has reported the GWTC-3 catalog of compact-binary coalescences from the O1, O2, and O3 observing runs \cite{Abbott2023GWTC3Catalog}. These observations make population inference a central tool for studying compact-object formation, black-hole mass spectra, spin and mass-ratio structure, and merger-rate evolution \cite{Zevin2021, Stevenson2017, Mandel2022, Abbott2023}. The scientific target is not only to fit the events already observed, but to infer the intrinsic population that generated a catalog after measurement uncertainty and selection effects have acted on it.

Future observatories will extend this program in complementary ways. The Laser Interferometer Space Antenna (LISA) \cite{Amaro-Seoane2017} will observe supermassive black hole binaries (SMBHBs) in the millihertz band, probing massive black-hole assembly over a wide redshift range. The expected number of LISA SMBHB detections is not generically enormous, but even modest catalogs can distinguish broad seed and growth scenarios if population-level selection effects are handled consistently. Other proposed space missions such as Taiji \cite{Hu2017} and TianQin \cite{Luo2016} would probe related frequency bands. On the ground, third-generation experiments such as the Einstein Telescope \cite{Punturo2010} and Cosmic Explorer \cite{Evans2021} are expected to substantially increase the number and redshift reach of stellar-mass compact-binary detections. These regimes differ in source populations, detector response, selection functions, and event-level measurement uncertainties, so a flexible population method must be validated in a detector-specific way before being transferred between instruments.

Inferring intrinsic source distributions from noisy and selection-biased catalogs is typically formulated as hierarchical Bayesian inference \cite{Mandel2019, Thrane2019, Vitale2020Selection}. This framework connects event-level posterior samples to population hyperparameters while accounting for parameter-estimation priors, measurement uncertainty, and the population-level detection efficiency \cite{Loredo2004, Mandel2019, Talbot2023, Mancarella2025}. The selection correction is essential because detected catalogs overrepresent sources that are easier to observe. For example, more massive or more favorably located systems can be visible over a larger spacetime volume, and a catalog of detections therefore does not directly trace the intrinsic population. A model-agnostic method must still respect this standard hierarchical structure: flexibility in the population density does not remove the need for a correct likelihood and a well-defined selection normalization.

Most current analyses rely on phenomenological parametric or semi-parametric models, such as power-law plus peak mass models, splines, binned reconstructions, Gaussian processes, or Dirichlet-process mixtures \cite{Abbott2023, Tiwari2021, Edelman2022, Callister2024, Rinaldi2024, Payne2025}. These methods have produced robust astrophysical constraints and remain the reference point for interpreting GW catalogs. Their strength is that the resulting parameters often have direct astrophysical meaning, and the priors can encode well-motivated physical assumptions. Their limitation is that any chosen model family can also impose structure on the inferred distribution. If the true population contains multiple subpopulations, sharp transitions, or correlations not represented by the adopted parameterization, a rigid model can smooth those features away or redistribute them into biased hyperparameters \cite{Fabbri2025, Payne2025, Edelman2022, Farah2023, Gennari2025}.

A model-agnostic framework is therefore useful as a complementary diagnostic alongside parametric analyses. Its role is to ask whether the catalog supports structure that is difficult to represent with a small number of fixed hyperparameters, while retaining the standard selection correction and posterior-prior reweighting. This is especially valuable when the physical interpretation is still being developed: for LISA SMBHBs, the relation between the observed mass-redshift distribution and seed formation channels is model dependent; for LVK binary black holes, reference population fits already exist, so a flexible reconstruction can be compared against them without turning the result into a complete astrophysical decomposition.

The appropriate sampler depends on the likelihood, parameterization, and validation task. Markov Chain Monte Carlo, nested sampling, and accelerated or machine-learning-assisted samplers are all widely used in GW inference, and each can be effective for suitable likelihoods and parameterizations \cite{Foreman-Mackey2013, Skilling2006, Talbot2018, Mandel2019}. In this work we develop a reusable amortized variational approach for flexible population densities, with explicit multistart stability criteria designed to identify unstable optima and to separate model selection from validation against external information.

Normalizing flows \cite{Rezende2015, Papamakarios2021, Lanchares2025, Colloms2025, Shih2024} and neural posterior-estimation techniques \cite{Dax2023, Jiang2025} provide a practical way to approximate complex catalog-level posteriors once the likelihood and selection correction have been specified. In this role, deep generative models complement existing samplers by providing reusable posterior approximations for repeated analyses, multistart validation campaigns, and sensitivity studies. In the framework developed here, the flow approximates the posterior over population hyperparameters, while the population density remains an explicit, evaluable distribution on a declared support.

Here we introduce a model-agnostic population-inference framework that couples a correlated compound-mixture population model to a normalizing-flow variational guide. The population density remains physically interpretable: finite mixtures describe multimodality, bounded marginal distributions enforce support constraints, and a Gaussian-copula structure captures inter-parameter dependence in a controlled way \cite{Adamcewicz2022, Adamcewicz2023}. This separation lets the method test joint structure without hiding the population model inside an implicit density estimator. We validate the framework in two regimes. First, controlled LISA SMBHB simulations provide known truth, allowing repeated truth-blind checks of shape and count recovery. Second, real LVK GWTC-3 binary black holes provide a detector-specific application using GWOSC posterior samples, official O1/O2/O3 injection assets, and official GWTC-3 reference curves.

The remainder of this paper is organized as follows. Sec.~\ref{sec:methodology} gives the hierarchical likelihood, the corrected selection treatment, the population model, and the truth-blind multistart stability criteria. Sec.~\ref{sec:lisa_application} presents the LISA simulation and repeated-validation results. Sec.~\ref{sec:lvk_application} presents the LVK GWTC-3 binary black hole application. Sec.~\ref{sec:conclusion} summarizes the supported conclusions. Throughout this work, we adopt a standard $\Lambda$CDM cosmological framework with $H_0 = 67.4 \, \text{km s}^{-1} \text{Mpc}^{-1}$ and $\Omega_m = 0.315$ \cite{Planck2020}.

\section{Methodology}
\label{sec:methodology}

Recovering the intrinsic astrophysical distribution constitutes a hierarchical inference problem \cite{Mandel2019, Thrane2019, Vitale2020Selection}. Our objective is to infer the posterior distribution for hyperparameters $\bm{\Lambda}$ that govern the underlying population, given event-level data $\{D_i\}$ and the detector selection function. Throughout this section, $\bm{\theta}$ denotes the source parameters used in a given application, for example $\{\log_{10}M_\bullet,z,q\}$ for the LISA validation or $\{\log_{10}m_1,z,q\}$ for the LVK analysis. This section presents the selection-corrected likelihood, the flexible population model, and the multistart stability criteria used in the detector-specific applications.

\subsection{Population Inference Framework}
\label{sec:hierarchical_framework}

An inhomogeneous Poisson process describes the detection of gravitational-wave events \cite{Mandel2019, Thrane2019, Vitale2020Selection}. For a population density $p_{\text{pop}}(\bm{\theta}|\bm{\Lambda})$ normalized over the intrinsic support, intrinsic annual rate $\mathcal{R}_{\text{yr}}$, observation time $T_{\text{obs}}$, and detection probability $P_{\text{det}}(\bm{\theta})$, the expected number of detected events is
\begin{equation}
\begin{split}
N_{\text{exp}}(\bm{\Lambda},\mathcal{R}_{\text{yr}})
&= \mathcal{R}_{\text{yr}} T_{\text{obs}} \alpha(\bm{\Lambda}),\\
\alpha(\bm{\Lambda})
&= \int P_{\text{det}}(\bm{\theta})
p_{\text{pop}}(\bm{\theta}|\bm{\Lambda}) d\bm{\theta}.
\end{split}
\label{eq:detection_efficiency}
\end{equation}
The detection efficiency $\alpha(\bm{\Lambda})$ is a population-level quantity: it is the average probability that an event drawn from the intrinsic population is detected. It depends on $\bm{\Lambda}$ because different population shapes place different fractions of events in the sensitive region of the detector.

For a catalog conditioned on detection, the selection process enters through the Poisson intensity in Eq.~\ref{eq:detection_efficiency}. The full likelihood for the measured event data and the observed count is therefore
\begin{equation}
\begin{split}
p(\{D_i\}, N_{\text{obs}} | \bm{\Lambda}, \mathcal{R}_{\text{yr}})
&\propto
e^{-\mathcal{R}_{\text{yr}}T_{\text{obs}}\alpha(\bm{\Lambda})}
\mathcal{R}_{\text{yr}}^{N_{\text{obs}}} \\
&\quad \times
\prod_{i=1}^{N_{\text{obs}}}
\int p(D_i|\bm{\theta}) p_{\text{pop}}(\bm{\theta}|\bm{\Lambda}) d\bm{\theta},
\end{split}
\label{eq:full_likelihood}
\end{equation}
up to constants independent of $\bm{\Lambda}$ and $\mathcal{R}_{\text{yr}}$. If the event-level parameter-estimation posterior was produced with prior $\pi_i(\bm{\theta})$, then
\begin{equation}
\int p(D_i|\bm{\theta}) p_{\text{pop}}(\bm{\theta}|\bm{\Lambda}) d\bm{\theta}
\propto
\left\langle
\frac{p_{\text{pop}}(\bm{\theta}_{ij}|\bm{\Lambda})}{\pi_i(\bm{\theta}_{ij})}
\right\rangle_{j \in p(\bm{\theta}|D_i)} .
\label{eq:event_reweighting}
\end{equation}
Eq.~\ref{eq:event_reweighting} is the posterior-prior reweighting used for both simulated posteriors and GWOSC posterior samples. The event terms depend on the intrinsic population density through this reweighting, while the selection correction is carried by $\alpha(\bm{\Lambda})$ and the Poisson normalization in Eq.~\ref{eq:full_likelihood}.

This separation gives the flexible density model a direct population interpretation. The event term measures how well an intrinsic population density explains the posterior samples for the detected events, and the selection term measures the detectable fraction of that same intrinsic population under the relevant search. The population-level normalization $\alpha(\bm{\Lambda})$ is evaluated with a differentiable detection model in the LISA validation and with recovered injections in the LVK analysis.

The population shape can be separated from the overall rate normalization by analytically marginalizing the rate scale with the usual scale-invariant count factor, equivalent to $p(\mathcal{R}_{\text{yr}})\propto 1/\mathcal{R}_{\text{yr}}$ for the count term. Dropping constants independent of $\bm{\Lambda}$ and writing $\mathcal{R}$ for $\mathcal{R}_{\text{yr}}$, this step gives
\begin{equation}
\begin{split}
\int_0^\infty d\mathcal{R}\,
\mathcal{R}^{N_{\text{obs}}-1}
e^{-\mathcal{R}T_{\text{obs}}\alpha(\bm{\Lambda})}
&=
\Gamma(N_{\text{obs}})
\left[T_{\text{obs}}\alpha(\bm{\Lambda})\right]^{-N_{\text{obs}}},
\end{split}
\end{equation}
and therefore the $\alpha(\bm{\Lambda})^{-N_{\text{obs}}}$ factor in the normalized-shape likelihood,
\begin{equation}
\label{eq:likelihood}
\begin{split}
p(\{D_i\} | \bm{\Lambda}) &\propto
\alpha(\bm{\Lambda})^{-N_{\text{obs}}}
\prod_{i=1}^{N_{\text{obs}}}
\left\langle
\frac{p_{\text{pop}}(\bm{\theta}_{ij}|\bm{\Lambda})}{\pi_i(\bm{\theta}_{ij})}
\right\rangle_j .
\end{split}
\end{equation}

The rate-marginalized likelihood in Eq.~\ref{eq:likelihood} is used for population-shape inference. After the selected LISA representatives have been chosen, the controlled validation also evaluates the implied expected counts. For that post-selection count-consistency calculation, the conditional posterior for the rate is
\begin{equation}
\begin{split}
p(\mathcal{R}_{\text{yr}} | N_{\text{obs}}, \bm{\Lambda})
&\propto p(\mathcal{R}_{\text{yr}})
\frac{e^{-\mathcal{R}_{\text{yr}} T_{\text{obs}} \alpha(\bm{\Lambda})}}
{N_{\text{obs}}!} \\
&\quad \times
\left[\mathcal{R}_{\text{yr}} T_{\text{obs}}
\alpha(\bm{\Lambda})\right]^{N_{\text{obs}}},
\end{split}
\label{eq:rate_posterior}
\end{equation}
where $p(\mathcal{R}_{\text{yr}})$ is the prior on the rate. In the controlled LISA count-validation study, we adopt the broad prior $\mathcal{R}_{\text{yr}}\sim \mathrm{LogNormal}[\log(N_{\text{obs}}/T_{\text{obs}}),2]$. For each posterior draw, the expected intrinsic and detected counts are
\begin{equation}
N_{\text{int}}^{\text{exp}}=\mathcal{R}_{\text{yr}}T_{\text{obs}},
\qquad
N_{\text{obs}}^{\text{exp}}=\mathcal{R}_{\text{yr}}T_{\text{obs}}\alpha(\bm{\Lambda}).
\end{equation}
These expected-count intervals are compared with the realized intrinsic and detected catalog counts only in the controlled LISA validation where the survey construction and truth are known. For real GWTC-3 data in Sec.~\ref{sec:lvk_application}, the reported posterior-predictive distributions are interpreted as catalog-level population shapes.

This hierarchical form provides a common route from a selection-biased detected catalog to an intrinsic population density. In controlled simulations, where the population truth and survey construction are known, it enables validation of both shape recovery and expected-count normalization. In real-data applications, it can be combined with injection-based selection correction and compared with established reference analyses.

\subsection{Correlated Compound Mixture Population Model}
\label{sec:network}

Astrophysical merger distributions can contain multimodality and parameter correlations \cite{Callister2024, Talbot2018}. Different formation channels may contribute to different regions of mass, redshift, spin, or mass ratio, and detector selection can then distort their apparent relative abundance. To capture such morphology without imposing a fixed parametric form, we construct a flexible population model $p_{\text{pop}}(\bm{\theta}|\bm{\Lambda})$ implemented as an explicit mixture density. This bounded and evaluable density enters the hierarchical likelihood directly, can be reweighted against posterior samples, and can be integrated against the selection function.

The first building block of our architecture is a finite mixture of $K$ distinct components, enabling the model to capture multimodal features that may correspond to different astrophysical formation channels \cite{Bishop2006}:
\begin{equation}
p(\bm{\theta} | \bm{\Lambda}) = \sum_{k=1}^{K} w_k \, p_k(\bm{\theta} | \bm{\lambda}_k).
\end{equation}
Here, $\{w_k\}$ denote the mixture weights satisfying $\sum_{k=1}^{K} w_k = 1$. This mixture model approximates the unknown astrophysical distribution as a weighted sum of simpler probability densities, providing a flexible basis for representing complex shapes. A small number of components can describe smooth unimodal populations, while additional components can represent shoulders, secondary peaks, or broader tails when the data support them. The number of components and component families are application-specific modeling choices. The LISA repeated-validation runs use a streamlined $K=2$ Kumaraswamy model, the LVK application uses the two-resolution axis-spline configuration described in Sec.~\ref{sec:lvk_inference}, and the toy validation in Sec.~\ref{sec:toy_model_validation} serves as a core density-estimation check.

The second building block addresses the modeling of correlations between parameters. We achieve this using a Gaussian copula \cite{Nelsen2006}, which separates the modeling of the marginal distributions from the dependence structure. Gaussian copulas capture elliptical correlation structures commonly observed in GW parameter dependencies \cite{Callister2024, Talbot2018} and provide tractable likelihood evaluation \cite{Chen2019}. Separating the marginals from the dependence structure allows controlled tests of correlation without confounding changes in the marginal distribution shapes. The joint probability density function for each mixture component is given by Sklar's theorem \cite{Sklar1959}:
\begin{equation}
\begin{split}
p_k(\theta_1, \dots, \theta_D | \bm{\lambda}_k)
&= c_k(F_{k,1}(\theta_1), \dots, F_{k,D}(\theta_D)) \\
&\quad \times \prod_{d=1}^{D} f_{k,d}(\theta_d).
\end{split}
\end{equation}
Here, $f_{k,d}$ and $F_{k,d}$ denote the marginal probability density and cumulative distribution functions respectively, and $c_k$ denotes the Gaussian copula density defined by a correlation matrix $\bm{R}_k$. To ensure positive definiteness during optimization, we parameterize $\bm{R}_k$ via its Cholesky decomposition. This separation of marginals and dependence supports controlled correlation tests, following the logic of recent copula-based catalog studies \cite{Adamcewicz2022, Adamcewicz2023}.

Explicitly, for $\bm{u}=(u_1,\ldots,u_D)$ with $u_d=F_{k,d}(\theta_d)$ and $\eta_d=\Phi^{-1}(u_d)$, the Gaussian copula density is
\begin{equation}
c_k(\bm{u})
=|\bm{R}_k|^{-1/2}
\exp\left[-\frac{1}{2}\bm{\eta}^{T}
\left(\bm{R}_k^{-1}-\bm{I}\right)\bm{\eta}\right],
\end{equation}
where $\Phi$ is the standard-normal cumulative distribution function. This expression makes the component density fully evaluable: the marginals set the one-dimensional shapes and supports, while $\bm{R}_k$ controls the dependence structure.

For the marginal densities $f_{k,d}$, the framework uses a library of flexible bounded forms. The library includes the Kumaraswamy, Truncated Pareto, and Truncated Normal distributions \cite{Kumaraswamy1980}. These distributions are useful because they can be normalized on the finite physical supports used in the analyses, such as bounded mass-ratio intervals and detector-specific mass or redshift ranges. The complete set of learnable hyperparameters $\bm{\Lambda}$ consists of the mixture weights and the parameters governing the marginals and copulas for all $K$ components.

We refer to this architecture as a Correlated Compound Mixture Density Network. The compound design preserves tractable density evaluation for the selection integral while providing enough flexibility to represent a broad class of non-Gaussian joint distributions without imposing a single prescriptive functional form.

\subsection{Amortized Variational Inference with Normalizing Flows}
\label{sec:VI}

Flexible population models can produce high-dimensional hyperposteriors with non-Gaussian structure. Direct sampling remains possible in principle, but repeated validation over many catalogs and random starts benefits from a posterior approximation that can be generated and evaluated quickly. We use stochastic variational inference (SVI) \cite{Hoffman2013} to approximate this posterior with a parametric guide $q_{\bm{\phi}}(\bm{\Lambda})$ by maximizing the evidence lower bound (ELBO) \cite{Blei2017}:
\begin{equation}
\mathcal{L}(\bm{\phi}) = \mathbb{E}_{q_{\bm{\phi}}(\bm{\Lambda})} \left[\log p(\{D_i\}|\bm{\Lambda})+\log p(\bm{\Lambda}) - \log q_{\bm{\phi}}(\bm{\Lambda})\right].
\end{equation}
Here $p(\{D_i\}|\bm{\Lambda})$ denotes the selection-corrected shape likelihood in Eq.~\ref{eq:likelihood}, and $p(\bm{\Lambda})$ is the hyperprior. Maximizing this bound minimizes the Kullback-Leibler divergence between the approximation and the true posterior probability density.

\begin{figure*}[htbp]
    \centering
    \includegraphics[width=0.9\textwidth]{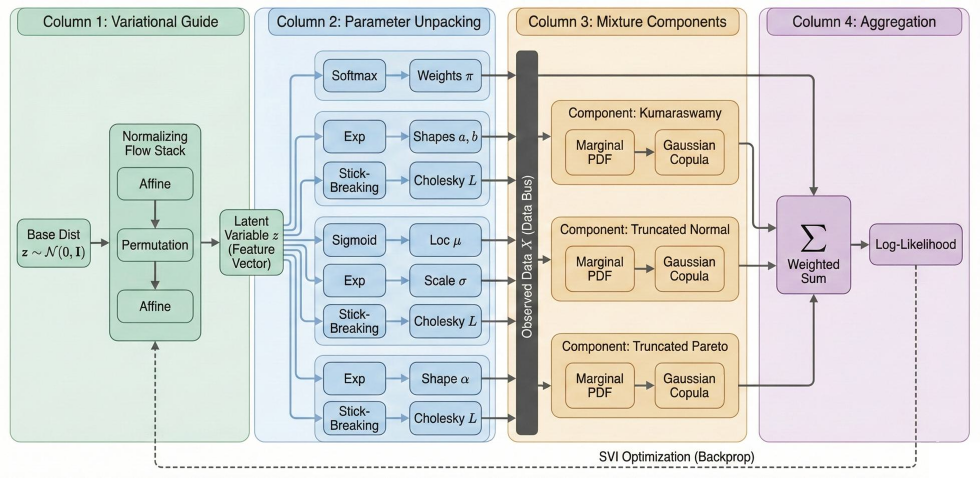}
    \caption{Flow-guided correlated compound-mixture population model. A latent variable $z$ is transformed by a normalizing-flow variational guide and unpacked into the parameters of the explicit mixture density. The mixture components use bounded marginal distributions, including Kumaraswamy, truncated normal, and truncated Pareto forms, together with copula-based dependence. The likelihood is evaluated by combining posterior-prior reweighting with a separate population-level selection correction. The figure emphasizes the separation between the variational guide for hyperparameters and the population density used in the hierarchical likelihood.}
    \label{fig:tech_architecture}
\end{figure*}

The fidelity of this approximation depends critically on the flexibility of the variational guide $q_{\bm{\phi}}$. Simple parametric families can miss non-Gaussian structure in the hyperposterior. We therefore employ a normalizing flow \cite{Papamakarios2021} as the guide. These models transform a simple base distribution such as a standard Gaussian into a more complex target density through a sequence of differentiable and invertible mappings. We use affine coupling layers \cite{Dinh2017}, which allow flexible transformations while keeping the Jacobian determinant efficient to evaluate. Fig.~\ref{fig:tech_architecture} depicts the architecture schematically, and Algorithm~\ref{alg:inference} details the inference procedure.

A useful feature of combining SVI with normalizing flows is amortized posterior approximation \cite{Mould2025, Dax2023}. Once trained for a specified model and data representation, the guide can generate hyperparameter samples rapidly, which is valuable for repeated validation runs and multistart stability studies. The realized speedup depends on the model, sampler, and numerical implementation, so we use this feature here as a practical computational property whose impact is assessed within each application. The flow guide samples $\bm{\Lambda}$, and each sampled $\bm{\Lambda}$ defines an explicit population density that can be evaluated in the likelihood and in the selection integral.

The numerical calculations use PyTorch \cite{paszke2019} and Pyro \cite{bingham2019}. The baseline variational guide uses affine coupling layers whose internal networks are multilayer perceptrons with ReLU activations. Application-specific model sizes, learning rates, supports, and training lengths are given in Secs.~\ref{sec:results_validation} and~\ref{sec:lvk_inference}.

\begin{algorithm}[H]
\caption{Flow-Guided Bayesian Population Inference}
\label{alg:inference}
\begin{algorithmic}[1]
\Require PE samples $\{\bm{\theta}_i^{(s)}\}_{i,s}$, PE priors $\pi_i^{(s)}$, selection samples or $P_\text{det}$, observation time $T_\text{obs}$
\Ensure Converged variational posterior $q_{\bm{\phi}}(\bm{\Lambda})$ and, when reported, expected-count intervals

\Statex \textbf{Phase 1: Shape Inference via SVI}
\State Initialize flow parameters $\bm{\phi}$
\While{ELBO not converged}
    \State Sample $\bm{\Lambda} \sim q_{\bm{\phi}}(\cdot)$ \Comment{Reparameterization trick}
    \State Evaluate population model $p_\text{pop}(\cdot|\bm{\Lambda})$ \Comment{Sec.~\ref{sec:network}}
    \For{$i = 1$ to $N_\text{obs}$}
        \State $\mathcal{L}_i \gets \frac{1}{N_\text{post}} \sum_{s} \frac{p_\text{pop}(\bm{\theta}_i^{(s)}|\bm{\Lambda})}{\pi_i^{(s)}}$ \Comment{Posterior-prior reweighting}
    \EndFor
    \State $\alpha(\bm{\Lambda}) \gets \int P_\text{det}(\bm{\theta}) \, p_\text{pop}(\bm{\theta}|\bm{\Lambda}) \, d\bm{\theta}$ \Comment{Population-level selection correction}
    \State $\log\mathcal{L} \gets \sum_i \ln \mathcal{L}_i - N_\text{obs} \ln \alpha(\bm{\Lambda})$ \Comment{Eq.~(\ref{eq:likelihood})}
    \State ELBO $\gets \log\mathcal{L} + \ln p(\bm{\Lambda}) - \ln q_{\bm{\phi}}(\bm{\Lambda})$
    \State $\bm{\phi} \gets \text{Adam}(\nabla_{\bm{\phi}} \text{ELBO})$
\EndWhile

\Statex \textbf{Phase 2: Optional Count-Normalization Inference}
\For{$m = 1$ to $M$} \Comment{$M$: number of posterior samples}
    \State Sample $\bm{\Lambda}^{(m)} \sim q_{\bm{\phi}}(\cdot)$
    \State Compute $\alpha^{(m)} \gets \alpha(\bm{\Lambda}^{(m)})$
    \State $p^{(m)}(\mathcal{R}_\text{yr}) \gets p(\mathcal{R}_\text{yr}|N_\text{obs},\bm{\Lambda}^{(m)})$ \Comment{Eq.~(\ref{eq:rate_posterior})}
\EndFor
\State \Return $q_{\bm{\phi}}(\bm{\Lambda})$ and, when used, $p(\mathcal{R}_\text{yr}|\text{data}) \approx \frac{1}{M} \sum_m p^{(m)}(\mathcal{R}_\text{yr})$
\end{algorithmic}
\end{algorithm}

\subsection{Truth-Blind Multistart Stability Criteria}
\label{sec:stability_criteria}

Because flexible density models can have multiple local optima, the validation and real-data studies use multistart stability criteria. These criteria are specified before validation comparisons are made. The purpose is to separate numerical stability from scientific interpretation: a result is retained only if independent starts find a reproducible solution under diagnostics that do not use withheld validation information. For the LISA repeated-validation study, each mock catalog is fitted from five independent training starts. Truth-blind catalog features define the stable cluster; the required minimum stable-cluster size is three starts. The selected medoid is chosen inside the stable cluster, with ties broken by the lowest training seed. Ground-truth count coverage and shape distances are reserved for validation after the medoid has been selected. For the LVK study, the corresponding one- and two-dimensional posterior-predictive summaries are evaluated on fixed grids; start-to-start distances between these summaries define the stable family.

For importance-sampling diagnostics, we monitor the effective sample size
\begin{equation}
N_{\text{eff}} = \frac{\left(\sum_j w_j\right)^2}{\sum_j w_j^2},
\end{equation}
where $w_j$ are the relevant posterior or injection weights. This numerical diagnostic follows the standard need to control Monte Carlo accuracy in posterior reweighting and empirical selection functions \cite{Farr2019SelectionAccuracy}; it identifies unstable starts and is kept separate from the astrophysical population model. For the LVK application, stability selection is based on posterior-predictive summaries and start-to-start distances.

\subsection{Validation of Core Density Estimation}
\label{sec:toy_model_validation}

We first validate the core density-estimation capability of the population model in a simplified toy setting, excluding selection effects and measurement uncertainty. For this validation, we employed a mixture model with $K=6$ components, comprising 2 Kumaraswamy, 2 Truncated Pareto, and 2 Truncated Normal distributions. This test checks whether the model can learn a bounded, non-Gaussian three-dimensional density from a sparse intrinsic sample; the detector-level validation is provided separately in Sec.~\ref{sec:lisa_application}.

\begin{figure}
    \centering
    \includegraphics[width=0.92\linewidth]{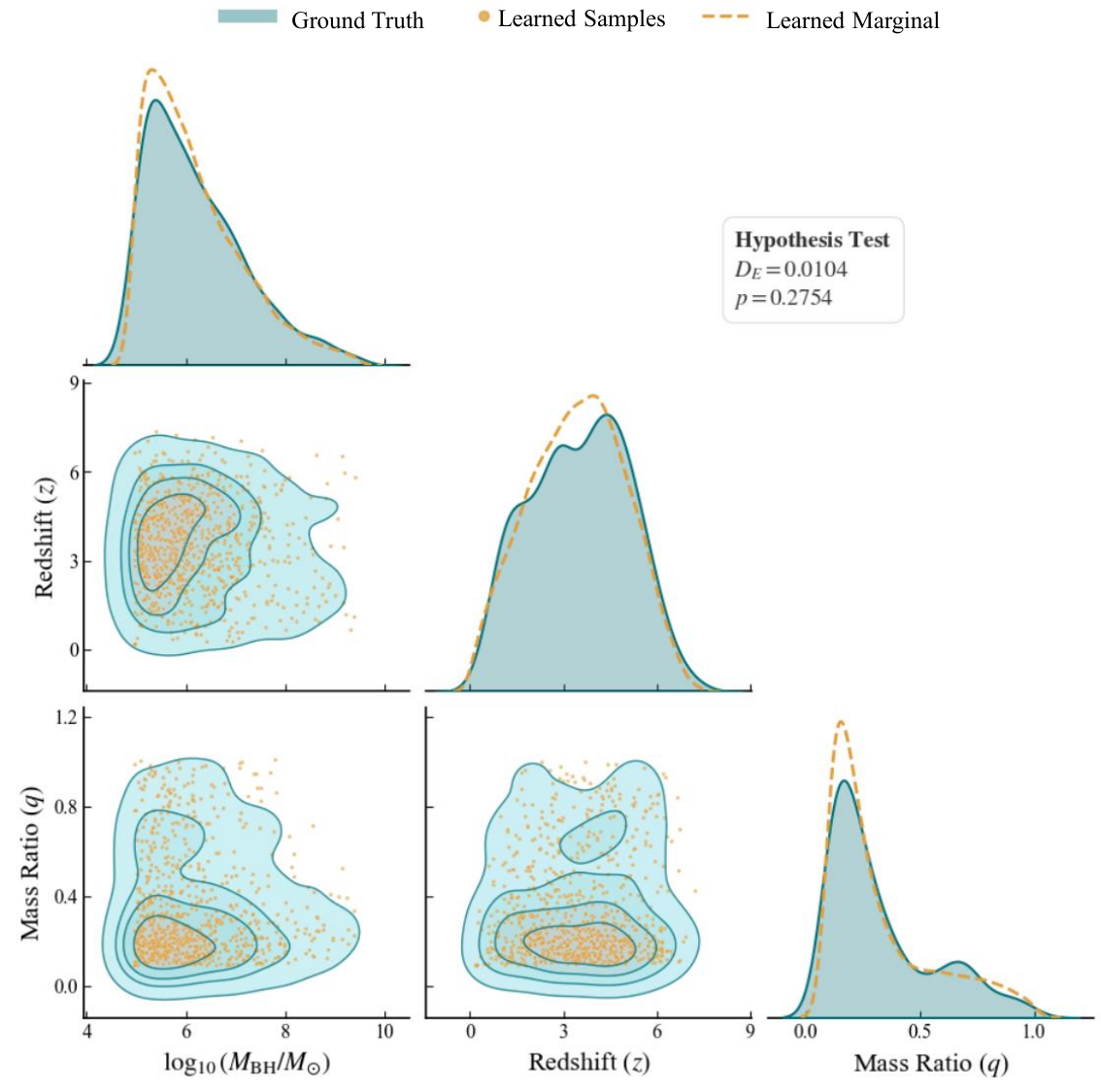}
    \caption{Validation of the density-estimation model on the toy problem without selection effects or measurement uncertainty. The corner plot compares the true intrinsic population (teal filled marginals and contours) with the learned samples and learned marginal densities (orange points and dashed curves), restricted to the bounded support used for the toy density. The inset reports the energy-distance statistic $D_E$ and its nonparametric test p-value.}
    \label{fig:toy_model_pairplot}
\end{figure}

We compared 10,000 samples from the learned distribution to the ground-truth catalog using a suite of statistical tests. This generated sample size reduces Monte Carlo noise in visual and distributional comparisons of the learned density. Kolmogorov-Smirnov (KS) tests on the marginal distributions yielded p-values of 0.274 for $\log_{10} M_{\bullet}$, 0.517 for $z$, and 0.259 for $q$, all above the standard significance threshold of 0.05. A nonparametric energy-distance test on the full three-dimensional joint distribution \cite{Szekely2013} yielded a p-value of 0.275. These tests do not reject consistency between the learned distribution and the ground truth in this simplified setting. Fig.~\ref{fig:toy_model_pairplot} shows the corresponding density comparison.

\section{Application to LISA: Supermassive Black Hole Binaries}
\label{sec:lisa_application}

The future LISA mission will offer a unique view of supermassive black hole binary mergers across cosmic time. For these sources, population inference is scientifically important because the mass-redshift distribution encodes information about black-hole seed formation, galaxy assembly, and the delay between galaxy mergers and black-hole coalescence. At the same time, the expected catalog size is small enough that validation cannot rely on asymptotic arguments alone. This section defines the controlled simulation used for validation and presents repeated truth-blind tests of shape and count recovery.

\subsection{Construction of the Intrinsic Population}
\label{sec:lisa_model}

To evaluate the performance of our inference framework, we first establish a physically motivated ground-truth population. Our intrinsic population model combines several key astrophysical ingredients to derive the comoving differential SMBH merger rate density, $\mathcal{R}(M_\bullet, q, z)$, where $M_\bullet$ denotes the primary SMBH mass and $q=M_{\bullet,2}/M_{\bullet,1}\leq1$. We begin with the galaxy stellar mass function, which describes the number density of host galaxies \cite{Weaver2023, Davidzon2017}, and convolve it with an empirical galaxy merger rate calibrated against cosmological simulations \cite{Rodriguez-Gomez2015, Fakhouri2010, Lacey1993, Springel2005}. Galaxies are connected to their central SMBHs via a scattered linear relation between the logarithmic black hole and stellar masses \cite{Kormendy2013, Reines2015, Pacucci2024, Volonteri2012}. Recently, similar analytical toy models have been explored to benchmark inference methods \cite{Langen2025}. This construction provides a realistic, nonuniform, and selection-affected validation target with known truth.

By marginalizing over all host galaxy properties, we obtain the intrinsic rate density. The corresponding differential event rate in the observer's frame is then
\begin{equation}
    \frac{d^3\dot{N}_{\text{int}}}{d\log_{10}M_{\bullet} \,dq \,dz} = \mathcal{R}(M_{\bullet}, q, z) \frac{dV_c}{dz} \frac{1}{1+z},
    \label{eq:observed_rate}
\end{equation}
where $V_c$ is the comoving volume element and the factor $(1+z)^{-1}$ converts source-frame time to observer-frame time. For a detailed list of all astrophysical parameters used, see Appendix~\ref{appendix:A}. The population model is defined over a three-dimensional parameter space: black hole mass $M_\bullet \in [10^{5}, 10^{10}]\,M_\odot$, or equivalently $\log_{10}(M_\bullet/M_\odot) \in [5, 10]$; redshift $z \in [0.21, 7.5]$; and mass ratio $q \in [0.1, 1.0]$. The bounded support defines the recovery domain for the validation and ties the inference target to the simulated survey. Integration of this rate density over a 4-year observation period yields an expected number of intrinsic events $N_{\text{int}} \approx 14$ and detectable events $N_{\text{obs}} \approx 13$. The repeated-validation catalogs below use controlled rescalings of this same intrinsic shape to test finite-sample behavior; the larger catalog sizes are validation stress tests and do not represent separate LISA rate forecasts.

\subsection{Simulation of the Observation Process}
\label{sec:lisa_observation}

For each intrinsic event, we model its observation by LISA. We employ the \texttt{IMRPhenomA} waveform model in the frequency domain to describe the GW signal from spinless binaries \cite{Ajith2007, Ajith2011}. Neglecting spin defines a spin-free controlled validation problem in the three population coordinates studied here. We quantify the detectability of each signal by its network signal-to-noise ratio, $\rho$, calculated using the LISA noise power spectral density \cite{Robson2018}, with instrumental noise and confusion noise from unresolved Galactic binaries \cite{Cornish2017}.

Detection is a probabilistic process that introduces selection bias. We model this bias with a smooth sigmoid function for the detection probability, $P_{\text{det}}(\rho)$, centered on a nominal SNR threshold of $\rho_{\text{th}}=8$ \cite{Klein2016} with a characteristic transition width of $\Delta \rho = 1.0$. We classify an intrinsic event as detectable based on a random draw weighted by this probability. To enable efficient gradient optimization, we employ a differentiable surrogate model for the SNR calculation \cite{Khan2021}. Implemented as a multilayer perceptron trained on precomputed SNR values, this surrogate makes repeated Monte Carlo selection-function evaluations practical for the validation study.

For each detected event, we simulate measurement uncertainty using the Fisher Information Matrix formalism \cite{Cutler1994, Vallisneri2008}. In the limit of high SNR, the posterior distribution over the parameters $\bm{\theta} = \{\log_{10}M_\bullet, z, q\}$ is well approximated by a multivariate Gaussian distribution, $\mathcal{N}(\bm{\theta}_{\text{true}}, \bm{\Gamma}^{-1})$, where $\bm{\Gamma}$ is the Fisher matrix. We draw posterior samples from this distribution for each event to emulate parameter-estimation posterior samples. Because the detection model is SNR based, the detected mock events preferentially occupy the regime where this Gaussian event-posterior approximation is most reliable.

\begin{figure*}[htbp]
    \centering
    \includegraphics[width=0.92\textwidth]{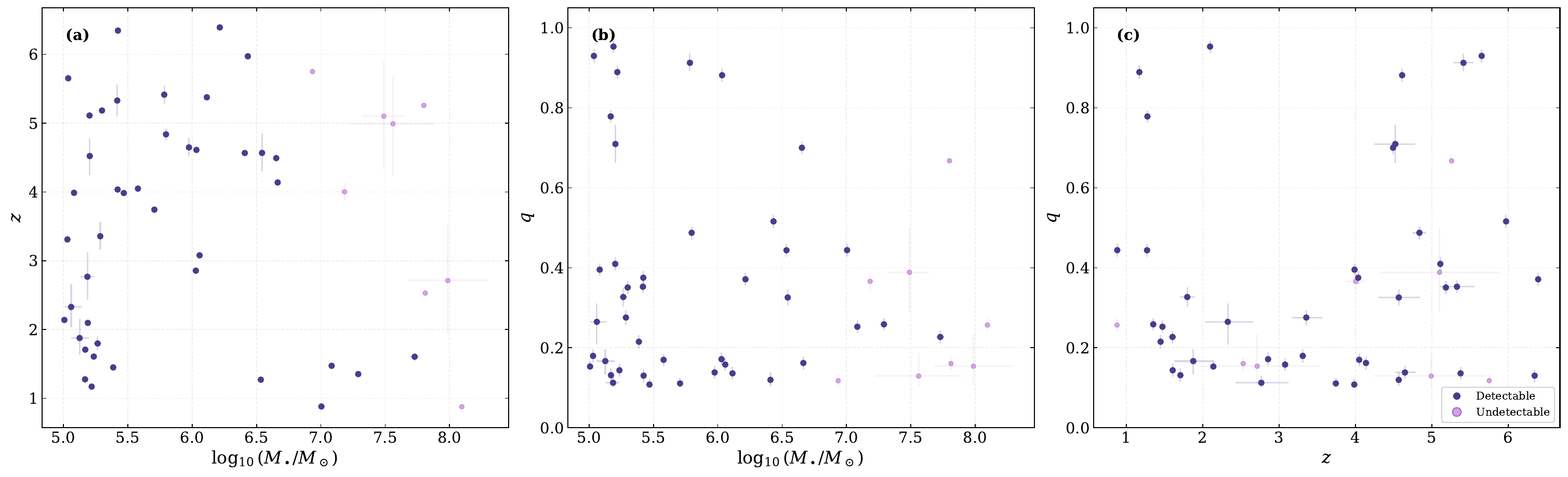}
    \caption{Projected distributions of the intrinsic ($N_{\text{int}}=50$) and detectable ($N_{\text{obs}}=42$) supermassive black hole binary merger populations in the LISA parameter space. Points show simulated source parameters; detectable events (dark slate blue) are distinguished from undetectable events (plum). \textbf{(a)} Primary SMBH mass versus redshift. \textbf{(b)} Primary SMBH mass versus mass ratio. \textbf{(c)} Redshift versus mass ratio. Error bars on detectable events represent 90\% Fisher-approximation credible intervals.}
    \label{fig:mock_data_3d}
\end{figure*}

To validate the robustness of the framework across catalog sizes relevant to LISA forecasts, we generated mock catalogs at multiple scales. Intrinsic populations of $N_{\text{int}} \in \{20, 50, 200, 500\}$ events span the regime from sparse LISA-like detections to optimistic or extended-mission yields. Fig.~\ref{fig:mock_data_3d} presents the projected distributions for a representative catalog with $N_{\text{int}}=50$ yielding $N_{\text{obs}}=42$, visualizing the intrinsic population and the detectable events, highlighting the selection bias and measurement uncertainties. The SNR values of these simulated events range from tens to tens of thousands. Extreme parameters, specifically low $q$ or high $\log_{10}M_\bullet$, yield lower SNR values and distinct error bars, whereas error bars are negligible for events with high SNR.
 
\subsection{Recovering the Intrinsic Distribution}
\label{sec:results_validation}

Having generated mock observations, we apply the framework to recover the intrinsic population over $\{\log_{10}M_\bullet,z,q\}$ and the expected-count normalization from biased observed catalogs. The primary challenge is to reconstruct a three-dimensional distribution from detected events whose sampling is distorted by the SNR-dependent selection function and whose coordinates are blurred by event-level posterior uncertainty. This recovery problem is the controlled analogue of the population inference that would be needed for LISA catalogs, but here the truth is known and can be used only after selection of the fitted model.

To address finite-realization variability directly, the validation uses repeated catalogs. We generated five independent catalogs at each $N_{\text{int}}\in\{20,50,200,500\}$ and fit each catalog from five independent initializations, giving 20 catalogs and 100 fits. The four catalog sizes span the regime from sparse LISA-like detections to optimistic or extended-mission yields. This design tests two effects simultaneously: whether the flexible density can recover the population shape as the number of events increases, and whether the expected-count normalization remains consistent after correcting for the selection function.

For these analyses, we used the streamlined model with $K=2$ Kumaraswamy components. The single-event posterior integrals use $N_{\text{post}}=1000$ Fisher-approximation posterior samples per detected event. The selection function is evaluated with the same differentiable LISA SNR surrogate used to generate the mock catalogs. The multistart stability rule is truth-blind: stable clusters are selected using training and posterior-predictive summaries only, while the true counts and shape distances are reserved for validation after selection. Thus the selected medoid for each catalog is chosen before inspecting whether it covers the true intrinsic count or minimizes the distance to the known population.

\begin{table}[tbp]
\centering
\caption{Truth-blind repeated-validation summary for the LISA mock catalogs. Each catalog size has five independent catalogs and five starts per catalog. The ``Stable cat.'' column counts catalogs whose selected representative passed the pre-specified stability screen. Coverage columns report the fraction of selected medoids whose 90\% expected-count intervals contain the true intrinsic count $N_{\text{int}}$ and observed count $N_{\text{obs}}$. The KS columns summarize the post-selection shape validation across $\log_{10}(M_\bullet/M_\odot)$, $z$, and $q$: $D_{\rm mean}$ is the largest coordinate-wise mean KS distance across the five selected medoids at fixed catalog size, and $D_{\rm max}$ is the largest single selected-medoid coordinate KS distance across those catalogs.}
\label{tab:lisa_qcv2_summary}
\scriptsize
\setlength{\tabcolsep}{2pt}
\begin{ruledtabular}
\begin{tabular}{@{}cccccccc@{}}
$N_{\rm int}$ & Cat. & Run & Stable cat. & $N_{\rm int}$ cov. & $N_{\rm obs}$ cov. & $D_{\rm mean}$ & $D_{\rm max}$ \\ \hline
20  & 5 & 25 & 5 & 1.0 & 1.0 & 0.1706 & 0.1965 \\
50  & 5 & 25 & 5 & 1.0 & 1.0 & 0.1384 & 0.1640 \\
200 & 5 & 25 & 5 & 1.0 & 1.0 & 0.0881 & 0.1165 \\
500 & 5 & 25 & 5 & 1.0 & 1.0 & 0.0704 & 0.0860 \\
\end{tabular}
\end{ruledtabular}
\end{table}

\begin{figure*}[htbp]
  \centering
  \includegraphics[width=0.92\textwidth]{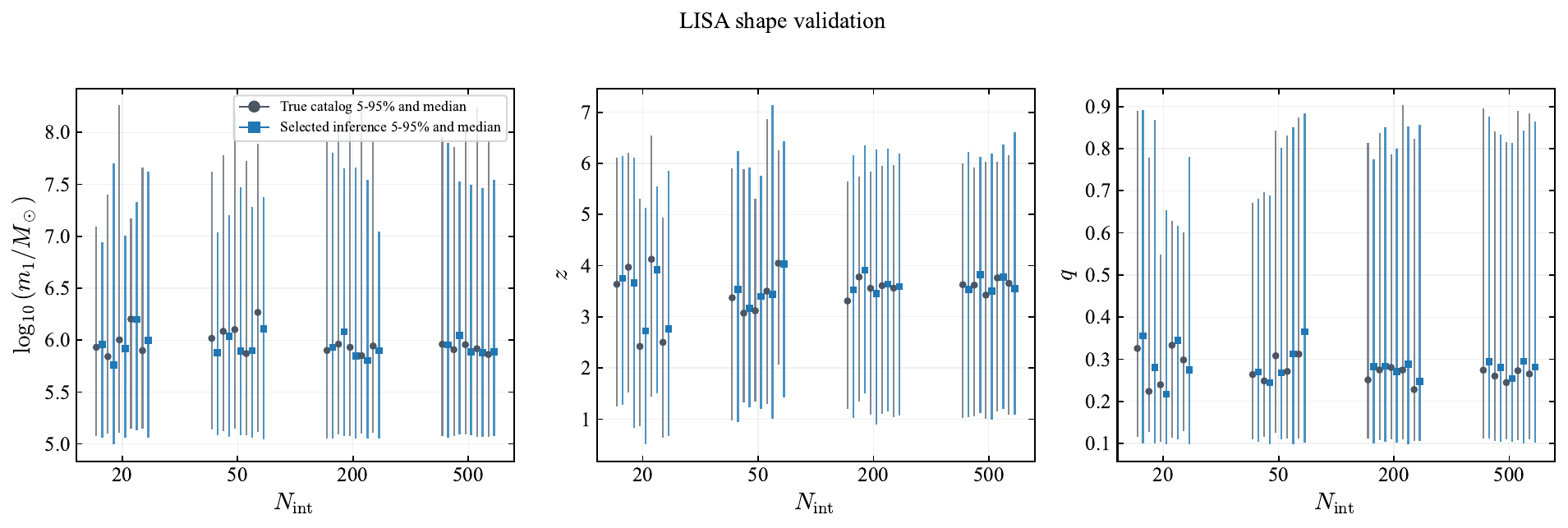}
  \caption{LISA shape-validation summary across 20 independent mock catalogs. At each catalog size, the five entries correspond to the five independent catalogs. Gray circles and intervals show the median and 5--95\% coordinate ranges of the true intrinsic catalog; blue squares and intervals show the corresponding summaries for the selected inferred intrinsic distribution. The comparison is made over the bounded LISA support $\log_{10}(M_\bullet/M_\odot)\in[5,10]$, $z\in[0.21,7.5]$, and $q\in[0.1,1]$; the KS summaries are reported in Table~\ref{tab:lisa_qcv2_summary}.}
  \label{fig:lisa_qcv2_shape_validation}
\end{figure*}

\begin{figure*}[htbp]
  \centering
  \includegraphics[width=0.88\textwidth]{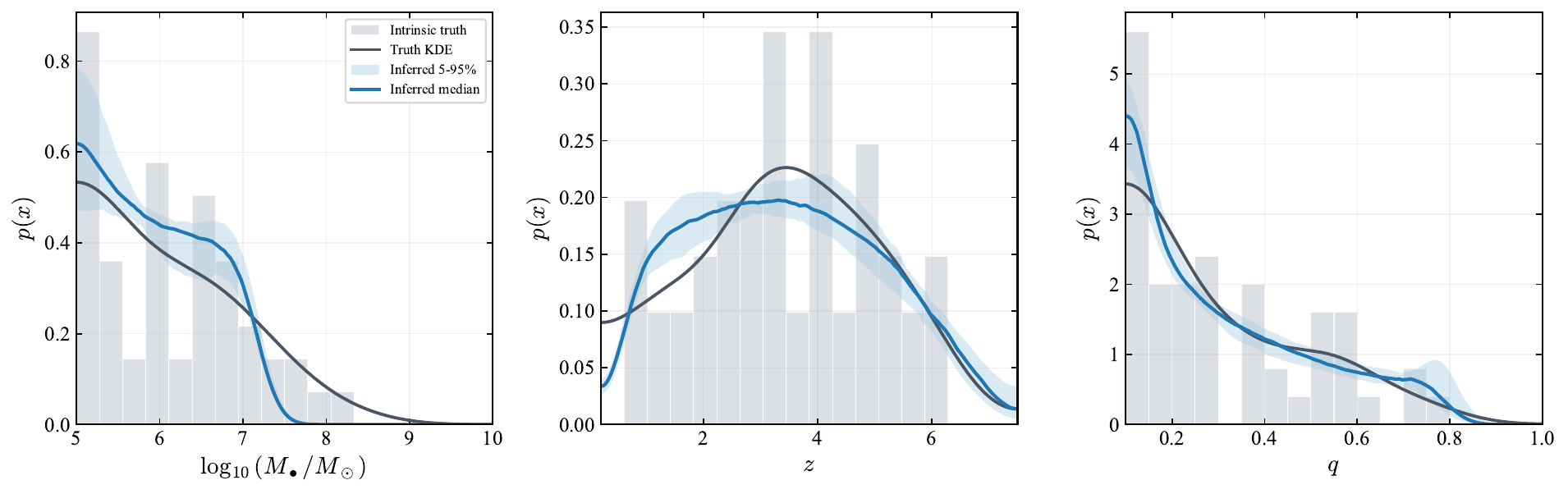}
  \caption{Representative LISA selected-medoid marginal-density comparison for the $N_{\text{int}}=50$ catalog shown in Fig.~\ref{fig:mock_data_3d}. The gray histograms and curves show the intrinsic simulation truth for this catalog. The blue band and blue curve show the selected-medoid inferred 5--95\% pointwise posterior-predictive marginal-density interval and median density, respectively. Local differences between the inferred marginals and the catalog-truth KDE reflect finite-realization sampling and event-level uncertainty for this representative catalog; the repeated-catalog KS trend in Table~\ref{tab:lisa_qcv2_summary} and Fig.~\ref{fig:lisa_qcv2_shape_validation} provides the quantitative shape validation.}
  \label{fig:lisa_qcv2_representative_marginals}
\end{figure*}

\begin{figure*}[htbp]
  \centering
  \includegraphics[width=0.90\textwidth]{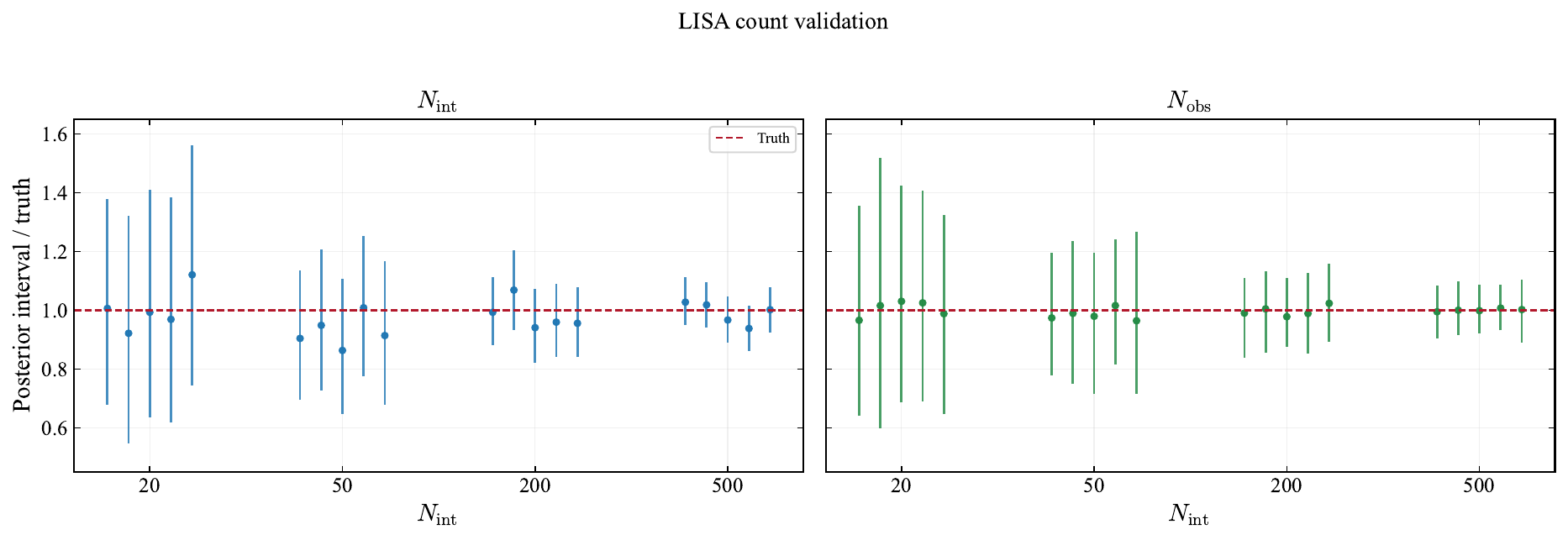}
  \caption{LISA count-validation summary. For each catalog size, the selected-medoid posterior intervals for the expected intrinsic and detected counts are divided by the true intrinsic count $N_{\text{int}}$ and the realized detected count $N_{\text{obs}}$, respectively. The dashed line at unity denotes exact agreement with the realized count; intervals crossing unity indicate coverage by the expected-count interval. All selected medoids pass the post-selection coverage test for both quantities.}
  \label{fig:lisa_qcv2_count_coverage}
\end{figure*}

Table~\ref{tab:lisa_qcv2_summary} summarizes the validation across all 20 catalogs and 100 independent initializations. For each catalog, the predefined stability criterion identifies a representative fit before the validation truth is inspected. All selected catalog-level representatives pass both intrinsic-count and observed-count expected-interval checks for every catalog size, with per-size coverage rates equal to 1.0. The maximum mean KS distance decreases from 0.1706 at $N_{\text{int}}=20$ to 0.0704 at $N_{\text{int}}=500$, consistent with improved shape recovery as the catalog grows. This trend is shown in Fig.~\ref{fig:lisa_qcv2_shape_validation}; Fig.~\ref{fig:lisa_qcv2_representative_marginals} gives a representative marginal-density comparison for the $N_{\rm int}=50$ catalog, and Fig.~\ref{fig:lisa_qcv2_count_coverage} displays the corresponding expected-count intervals across all catalog sizes.

The coverage fractions should be interpreted with finite-binomial uncertainty because there are five catalogs per size: a 5/5 success rate has an approximate 95\% Wilson interval of $0.57$--$1.00$, while the pooled 20/20 result has an interval of $0.84$--$1.00$; if the true coverage were 0.9, observing 20/20 successful checks would have probability about 0.12. The average relative widths of the 90\% expected-count intervals for $(N_{\text{int}},N_{\text{obs}})$ are approximately $(0.77,0.75)$, $(0.47,0.47)$, $(0.25,0.27)$, and $(0.16,0.18)$ for $N_{\text{int}}=20,50,200,$ and 500, respectively. The largest residual discrepancies occur in the sparse-catalog regime, as expected when a small number of detected events must constrain a three-dimensional intrinsic distribution. These discrepancies remain within the count-coverage behavior summarized in the table.

These repeated-validation results show that, under the specified LISA simulation, Fisher-posterior approximation, and streamlined population model, the inference framework recovers population shape and gives expected-count normalization consistent with the realized counts over the tested catalog sizes. The controlled setting also tests the combined action of posterior reweighting, selection correction, multistart stability selection, and expected-count normalization when the data-generating process is known.

The representative catalog visualization in Fig.~\ref{fig:mock_data_3d} provides the physical context for this validation. It shows how the detection process removes part of the intrinsic distribution and how event-level uncertainty varies across the LISA parameter space. The repeated-validation figures and table quantify the resulting recovery over independent catalog realizations.

\subsection{Comparison with a Parametric Estimation Method}
\label{sec:comparison}

As a focused flexibility check, we compare the model with a parametric estimator in simplified LISA-like tests. In a two-dimensional setting with $q=1$ fixed, a correctly specified parametric model gives the expected near-exact reconstruction. The flexible model remains competitive in this setting, with a Pearson correlation of 0.94 against the truth. In a deliberately misspecified test with a bimodal mass distribution and nonmonotonic redshift evolution, the flexible model better follows the injected morphology, with a Jensen--Shannon divergence of 0.11 compared with 0.22 for the restricted parametric form. These results support the role of the method as a model-agnostic complement to parametric analyses within this controlled comparison. The detailed diagnostic tables are given in Appendix~\ref{appendix:parametric_comparison}.

\section{Application to LVK: GWTC-3 Binary Black Holes}
\label{sec:lvk_application}

The LISA study provides a controlled validation against known truth. We next apply the framework to real stellar-mass binary black hole (BBH) events from the LVK GWTC-3 era. In this application, the event-level posteriors come from full LVK parameter-estimation products, the selection function is represented by recovered search injections, and the true astrophysical population is unknown. We therefore compare the inferred population shape with official GWTC-3 reference curves using detector-specific posterior samples and injection assets.

\subsection{Data and Selection Effects}
\label{sec:lvk_data_selection}

We use GWOSC posterior samples for confident GWTC-3-era events associated with the O1, O2, and O3 observing runs \cite{Abbott2023GWTC3Catalog, LVK2021GWTC3PEData}. The starting set consists of 89 deduplicated confident GWOSC parameter-estimation files from the GWTC-3-era releases. The retained sample is defined by the availability of a posterior group containing source-frame masses, redshift, mass ratio, and usable \texttt{log\_prior} values under the declared BBH support required for Eq.~\ref{eq:event_reweighting}. The event selection is fixed before fitting, leaving 75 retained events.

\begin{figure*}[htbp]
    \centering
    \includegraphics[width=0.92\textwidth]{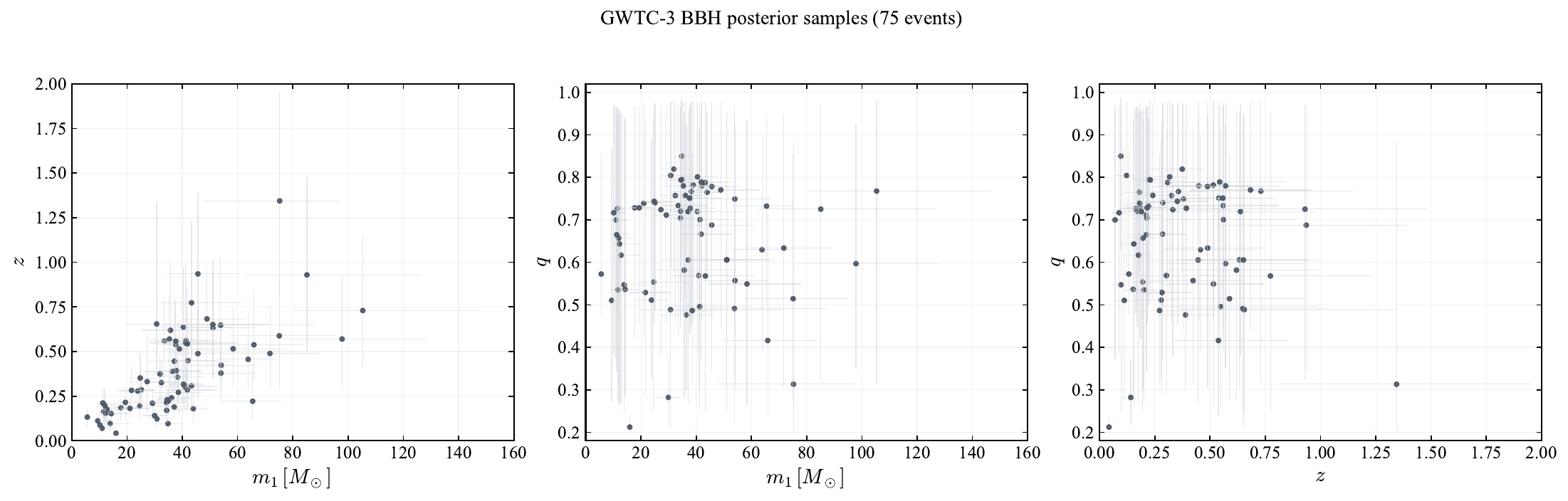}
    \caption{Overview of the 75 retained GWTC-3-era BBH events used in the LVK morphology comparison. The panels show posterior medians and 90\% credible intervals for \textbf{(a)} primary mass versus redshift, \textbf{(b)} primary mass versus mass ratio, and \textbf{(c)} redshift versus mass ratio. The event-display panels show $0\le z\le2$ and $q\in[0.2,1]$, with the primary-mass axes covering the retained-event range over the declared mass support. The inference support is $m_1\in[3,150]\,M_\odot$, $z\in[0,3]$, and $q\in[0.2,1]$.}
    \label{fig:lvk_gwtc3_events}
\end{figure*}

For each retained event, we use 5000 posterior samples in source-frame primary mass $m_1$, redshift $z$, and mass ratio $q=m_2/m_1$. Current ground-based BBH detections often have finite signal-to-noise ratios that leave broad event-level posteriors, especially in redshift and mass ratio. The LVK application therefore keeps the posterior-sample representation in the hierarchical reweighting step, so that measurement uncertainty is propagated into the catalog-level population density. The population support is $m_1\in[3,150]\,M_\odot$, $z\in[0,3]$, and $q\in[0.2,1]$. These boundaries define the analysis domain on which posterior reweighting and injection-based selection evaluation are numerically stable. The lower mass-ratio boundary defines the analysis region in which the posterior-prior and injection reweighting remain stable for the retained GWTC-3 samples. Similar finite-\(q\) restrictions appear in flexible GWTC-3 population studies when avoiding prior-dominated regions, and autoregressive comparisons with the GWTC-3 PowerLaw+Peak model identify an effective low-\(q\) truncation near \(q\simeq0.2\) from the \(m_{\min}/m_1\) support condition \cite{Heinzel2025PixelPop,Callister2024,Abbott2023}. Fig.~\ref{fig:lvk_gwtc3_events} shows the retained events and their event-level uncertainties.

Unlike the analytical SNR approximation used for LISA, the selection function for ground-based searches is characterized empirically through injection campaigns \cite{Farr2019SelectionAccuracy}. This is necessary because the sensitivity of the search depends on the detector network, observing-run duty cycles, nonstationary noise, and search recovery criteria. We use the official O1/O2/O3 injection file with the O3 any-IFAR recovery mask \cite{LVK2023GWTC3SensitivityData}. After applying the analysis support cuts, 40,286 recovered injections remain for selection-function evaluation. The detection efficiency is computed with importance sampling over recovered injections:
\begin{equation}
    \alpha(\bm{\Lambda}) = \frac{1}{N_{\text{draw}}} \sum_{j \in \text{det}} \frac{p_{\text{pop}}(\bm{\theta}_j | \bm{\Lambda})}{p_{\text{draw}}(\bm{\theta}_j)},
\end{equation}
where the sum runs over recovered injections, $p_{\text{draw}}$ is the injection sampling distribution, and $N_{\text{draw}}$ is the total number of generated injections in the campaign. This empirical normalization lets the same population density be evaluated against the actual search sensitivity represented by the injection campaign.

All LVK posterior-prior and injection weights are evaluated in the same source-frame coordinate measure, $\bm{x}=(\log_{10}m_1,z,q)$. The GWOSC prior information and the injection sampling density are therefore transformed into this measure before reweighting. For a density originally expressed in source-frame masses and redshift, the change from $(m_1,m_2,z)$ to $(\log_{10}m_1,q,z)$ contributes the Jacobian
\begin{equation}
J=\left|\frac{\partial(m_1,m_2,z)}{\partial(\log_{10}m_1,q,z)}\right|
=\ln(10)m_1^2 .
\end{equation}
The population density, the PE prior density, and $p_{\text{draw}}$ in the LVK likelihood are thus compared in the same coordinate measure.

\subsection{Inference Methodology}
\label{sec:lvk_inference}

The posterior-prior reweighting in Eq.~\ref{eq:event_reweighting} is applied to the GWOSC posterior samples using the provided prior information:
\begin{equation}
    w_{is} \propto \frac{p_{\text{pop}}(\bm{\theta}_{is} | \bm{\Lambda})}{\pi_i(\bm{\theta}_{is})},
\end{equation}
where $s$ indexes posterior samples for event $i$ and $\pi_i$ is the prior density used in parameter estimation after the coordinate transformation described above. This step removes the event-level parameter-estimation prior before applying the population model, allowing the same hierarchical likelihood used in the LISA validation to be applied to real posterior samples.

The LVK population density uses a two-resolution basis in $\log_{10}(m_1/M_\odot)$, with 5 global bins and 16 local bins over $5$--$50\,M_\odot$. This basis preserves a broad support for the high-mass tail while giving the lower-mass region enough resolution to compare with the structured mass spectra commonly used in GWTC-3 population analyses. The axis-spline mixture has 4 Kumaraswamy and 2 truncated-normal components. The variational guide uses 6 flow transforms with hidden dimension 128. We train for 1500 epochs after a 500-epoch warmup using Adam with learning rate $5\times10^{-4}$ and gradient clipping at norm 5.0.

The final LVK analysis uses a multistart stability selection based on posterior-predictive shape summaries and start-to-start distances. The reference curves shown below are taken from the official GWTC-3 population analysis and data release \cite{Abbott2023, LVK2024GWTC3PopulationData}.

\subsection{Inferred Population Distributions}
\label{sec:lvk_results}

\begin{figure*}[htbp]
    \centering
    \includegraphics[width=0.86\textwidth]{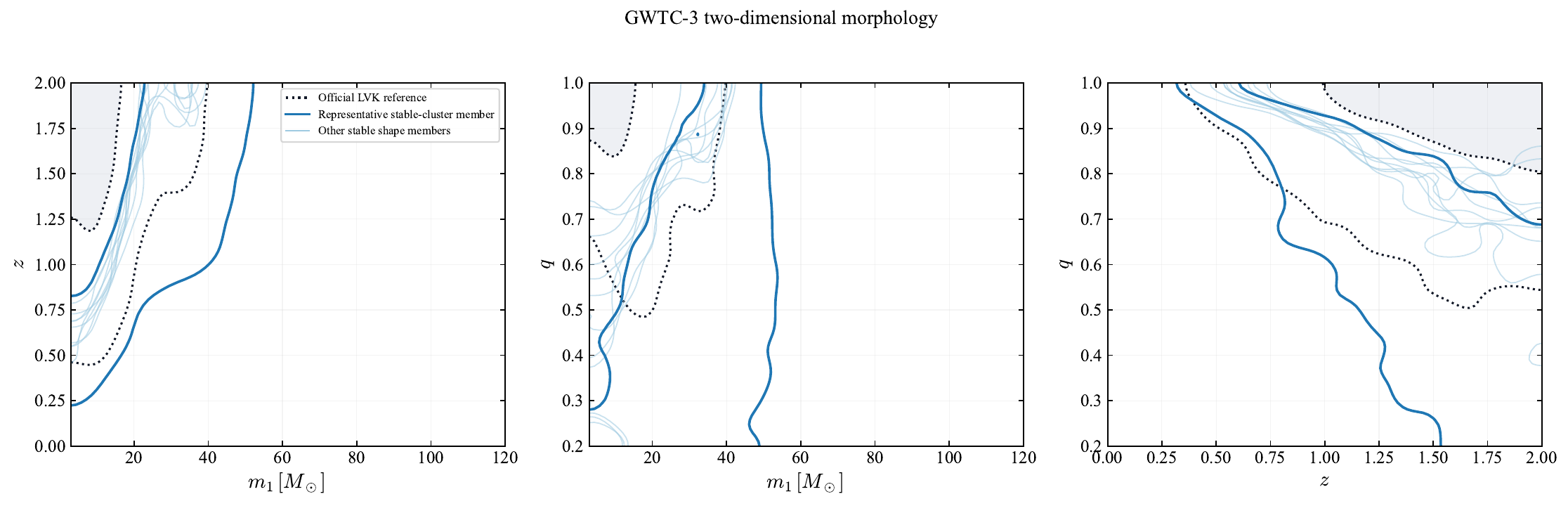}
    \caption{Two-dimensional normalized posterior-predictive morphology for the LVK GWTC-3 BBH analysis. The panels show \textbf{(a)} primary mass and redshift, \textbf{(b)} primary mass and mass ratio, and \textbf{(c)} redshift and mass ratio. Light-blue contours show the 50\% highest-density contour for other stable-family members; the solid blue curves show the 50\% and 90\% highest-density contours for the representative stable-cluster member. Dotted black contours show the corresponding 50\% and 90\% contours from the official GWTC-3 reference samples. The displayed window is $m_1\le120\,M_\odot$ and $z\le2$, while the inference support remains that listed in Table~\ref{tab:lvk_morphology_conventions}.}
    \label{fig:lvk_stage3p_distribution_2d}
    \vspace{0.3em}
    \includegraphics[width=0.86\textwidth]{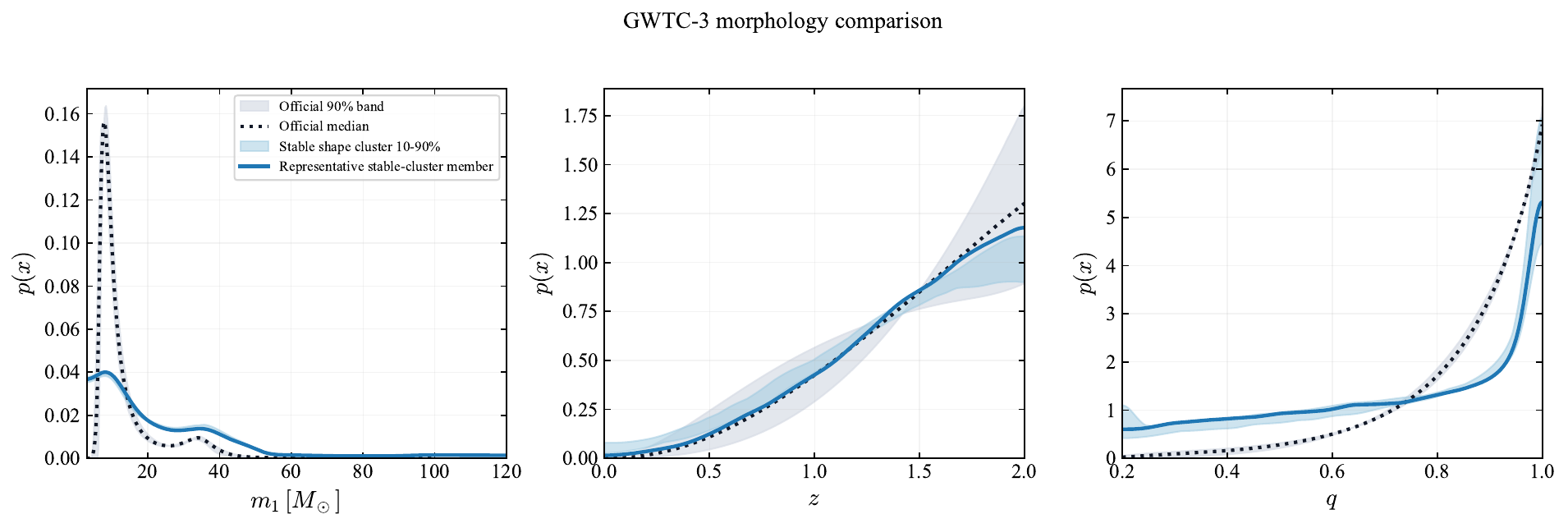}
    \caption{Normalized marginal morphology for the LVK GWTC-3 BBH analysis. The stable-cluster envelope is compared with official GWTC-3 reference curves for external context. The solid curve is a representative stable-cluster member. The primary-mass panel is displayed over $m_1\le120\,M_\odot$, and the redshift panel is conditionally normalized over $0\leq z\leq2$ for visualization; the inference support remains $m_1\le150\,M_\odot$ and $0\leq z\leq3$.}
    \label{fig:lvk_stage3p_marginal}
\end{figure*}

Fig.~\ref{fig:lvk_stage3p_marginal} compares the inferred one-dimensional marginals with official GWTC-3 reference curves. Fig.~\ref{fig:lvk_stage3p_distribution_2d} shows the corresponding two-dimensional structure. The stable-cluster envelope shows the start-to-start variation within the truth-blind stable family, while the solid line gives one representative stable-cluster member for visual comparison. Table~\ref{tab:lvk_morphology_conventions} records the data sources, support, coordinate measure, and display convention used for this comparison.

\begin{table*}[t]
\centering
\caption{LVK morphology-comparison conventions. The table records the data and interpretation boundaries for the GWTC-3 BBH application.}
\label{tab:lvk_morphology_conventions}
\begin{ruledtabular}
\begin{tabular}{@{}ll@{}}
Quantity & \multicolumn{1}{c}{Convention} \\ \hline
Event data & \begin{minipage}[t]{0.66\textwidth}GWOSC GWTC-3-era confident BBH posterior samples\end{minipage} \\
Selection data & \begin{minipage}[t]{0.66\textwidth}Official O1/O2/O3 injection assets with O3 any-IFAR recovery mask\end{minipage} \\
Analysis support & \begin{minipage}[t]{0.66\textwidth}$m_1\in[3,150]\,M_\odot$, $z\in[0,3]$, $q\in[0.2,1]$\end{minipage} \\
Coordinate measure & \begin{minipage}[t]{0.66\textwidth}Source-frame $(\log_{10}m_1,z,q)$ with prior/draw-density Jacobian included\end{minipage} \\
Official reference curves & \begin{minipage}[t]{0.66\textwidth}Post-selection external morphology comparison and overlay only\end{minipage} \\
Redshift visualization window & \begin{minipage}[t]{0.66\textwidth}Conditional $0\leq z\leq2$; inference support remains $0\leq z\leq3$\end{minipage} \\
Reported LVK quantity & \begin{minipage}[t]{0.66\textwidth}Normalized morphology\end{minipage} \\
\end{tabular}
\end{ruledtabular}
\end{table*}

The primary-mass marginal is shown directly in physical mass units, which facilitates comparison with published GWTC-3 population analyses. The inferred stable-cluster reconstruction captures a concentration at lower stellar-mass black-hole masses and a decline toward the upper end of the support, but the reconstructed low-mass structure is smoother and broader than the official reference median. This behavior indicates that the flexible density captures the broad catalog structure without resolving the sharper low-mass feature as a specific formation signature.

The mass-ratio and two-dimensional panels show the correlations and support regions favored by the flexible density after posterior-prior reweighting and injection-based selection correction. As a quantitative comparison, the one-dimensional Jensen--Shannon distances between the representative stable-cluster density and the official GWTC-3 reference marginals are $D_{\rm JS}(m_1)=0.111$, $D_{\rm JS}(z)=0.004$, and $D_{\rm JS}(q)=0.075$, giving a maximum one-dimensional distance of 0.111 under the same source-frame measure and normalization used in the figures.

The redshift marginal in Fig.~\ref{fig:lvk_stage3p_marginal} is displayed as a conditional density over $0\leq z\leq2$ for visual comparison with the official GWTC-3 reference curves, and the primary-mass display is restricted to $m_1\le120\,M_\odot$. The fitted population support used in the inference remains $m_1\le150\,M_\odot$ and $0\leq z\leq3$. The high-redshift tail of the GWTC-3 BBH sample is sparsely populated, and the flexible density does not impose a parametric redshift-evolution law. The $z\le2$ display window therefore emphasizes the better-constrained part of the catalog while keeping the inference domain fixed. Overall, the LVK application demonstrates that the same selection-corrected, model-agnostic framework can be configured for a ground-based catalog under the stated support, data cuts, and stability criteria.

\section{Conclusion}
\label{sec:conclusion}

We have presented a model-agnostic hierarchical population-inference framework that combines an explicit correlated compound-mixture density with a normalizing-flow variational guide. The population model remains an evaluable density with bounded marginals, mixture components, and copula-based dependence; the flow is used to approximate the posterior over the population hyperparameters. This separation is useful because it preserves the standard hierarchical likelihood structure while allowing the morphology of the population to be more flexible than a small set of fixed parametric forms. The controlled parametric comparison in Appendix~\ref{appendix:parametric_comparison} illustrates the corresponding tradeoff: correctly specified parametric forms remain efficient, while the flexible density reduces dependence on a prespecified functional form in the misspecified test.

A central technical point is the selection correction. Selection enters through $\alpha(\bm{\Lambda})$ and the Poisson normalization, while event information enters through the posterior-prior reweighted marginal likelihood. The same normalized population density $p_{\rm pop}(\bm{\theta}|\bm{\Lambda})$ is therefore used consistently in event reweighting and in selection-function integration.

For simulated LISA supermassive black hole binary catalogs, truth-blind repeated validation across 20 catalogs and 100 independent initializations supports shape recovery and expected-count normalization in the controlled simulation described here. All catalogs satisfy the stability criterion, the selected medoids' expected-count intervals cover the intrinsic and observed counts, and the shape discrepancies decrease with catalog size over the tested range. These results show that multistart stability selection, posterior reweighting, selection correction, and expected-count normalization operate coherently in the controlled LISA setting.

For LVK GWTC-3 binary black holes, we configure the framework with GWOSC posterior samples, official O1/O2/O3 injection assets, and detector-specific support cuts. The inferred distribution is broadly compatible with the official GWTC-3 reference curves while remaining smoother in the low-mass region than the reference median. This real-data application complements established population studies by checking whether a model-agnostic density can recover the main catalog-level structure under truth-blind stability criteria, while detailed astrophysical decomposition remains the role of analyses built for that purpose.

Many applications of the framework remain open. Future work will extend the validation to additional source parameters, including spin, and to broader detector configurations. For LISA, a natural next step is to connect recovered mass-redshift morphology to explicit SMBH seed and growth models after validating the inference on a wider class of simulations. For ground-based catalogs, future extensions should include spins, alternative support choices, and direct comparisons among flexible-density families under the same stability criteria. The main conclusion of the present study is that model-agnostic density models can be combined with standard hierarchical selection correction and truth-blind multistart stability selection to provide a practical complement to parametric gravitational-wave population analyses.

\section*{Data and Code Availability}

The reproducibility repository containing the analysis code, configuration files, compact derived summaries, and manuscript figure assets is available at \href{https://github.com/l46430640-del/Model-Agnostic-GW-Population-Inference-Reproducibility-Snapshot}{\texttt{https://github.com/\allowbreak{}l46430640-del/\allowbreak{}Model-Agnostic-GW-\allowbreak{}Population-Inference-\allowbreak{}Reproducibility-Snapshot}}. The repository does not redistribute large official LVK data products; instead, it provides download manifests and source links for the public GWOSC, O1/O2/O3 injection, and GWTC-3 population-release data used in the LVK application.

\begin{acknowledgments}
This work was supported by the National Key R\&D Program of China (grant No. 2024YFA1611503), the Central Guidance for Local Science and Technology Development Fund (grant No. ZYYD2026JD01), and the Urumqi Nanshan Astronomy and Deep Space Exploration Observation and Research Station of Xinjiang (XJYWZ2303). 

YF was supported by the National Natural Science Foundation of China (Grant No. 12405068).

\end{acknowledgments}

\appendix

\section{LISA Mock Data Generation Details}
\label{appendix:A}

This appendix details the specific parameters and mathematical justifications for the mock data catalog described in Sec.~\ref{sec:lisa_application}.

\subsection{Astrophysical Model Parameters}

We assume a $\Lambda$CDM cosmology with $H_0 = 67.4 \, \text{km s}^{-1} \text{Mpc}^{-1}$, $\Omega_m = 0.315$, and $\Omega_\Lambda = 0.685$ \cite{Planck2020}.
The Galaxy Stellar Mass Function (GSMF) uses a double Schechter function with redshift-dependent parameters interpolated from Table~\ref{tab:gsmf_params} (based on COSMOS2020 \cite{Weaver2023}).

\begin{table}[t]
\centering
\caption{Discrete data points for the redshift evolution of the GSMF parameters, from \cite{Weaver2023}.}
\label{tab:gsmf_params}
\begin{ruledtabular}
\begin{tabular}{cccccc}
\textbf{$z$} & \textbf{$\log_{10}(M_c/M_\odot)$} & \textbf{$\log_{10}(\phi_1^*)$} & \textbf{$\alpha_1$} & \textbf{$\log_{10}(\phi_2^*)$} & \textbf{$\alpha_2$} \\ \hline
0.35 & 10.93 & -3.22 & -1.45 & -2.97 & -0.66 \\
0.65 & 10.98 & -3.32 & -1.44 & -3.04 & -0.79 \\
0.95 & 11.01 & -3.15 & -1.35 & -3.11 & -0.72 \\
1.30 & 11.01 & -3.16 & -1.34 & -3.49 & -0.56 \\
1.75 & 10.88 & -3.64 & -1.52 & -3.17 & -0.55 \\
2.25 & 10.80 & -3.59 & -1.46 & -3.54 & 0.04 \\
2.75 & 11.00 & -3.66 & -1.46 & -9.00 & 0.00 \\
3.25 & 10.85 & -3.70 & -1.46 & -9.00 & 0.00 \\
4.00 & 10.46 & -3.70 & -1.46 & -9.00 & 0.00 \\
5.00 & 10.33 & -3.89 & -1.46 & -9.00 & 0.00 \\
6.00 & 10.19 & -4.30 & -1.46 & -9.00 & 0.00 \\
7.00 & 10.50 & -5.00 & -1.46 & -9.00 & 0.00 \\
\end{tabular}
\end{ruledtabular}
\footnotetext{Note: The units for $\phi_{1,2}^*$ are $\text{Mpc}^{-3}$.}
\end{table}

The $M_{\bullet}-M_{\star}$ relation transitions via a sigmoid at $z=4.0$ from a low-redshift regime ($b=1.17, a=-4.18, \epsilon=0.28$ \cite{Kormendy2013}) to a high-redshift regime ($b=1.06, a=-2.43, \epsilon=0.69$ \cite{Pacucci2024}). The galaxy merger rate follows \cite{Rodriguez-Gomez2015}.

We approximate the black hole mass ratio as the stellar mass ratio, $q \approx \mu_*$. Assuming the mean log-black hole mass follows $\langle\log_{10}M_{\bullet}\rangle = a(z) + b(z)\log_{10}M_*$ with Gaussian scatter $\epsilon(z)$, the mass ratio $q = M_{\bullet,2} / M_{\bullet,1}$ satisfies:
\begin{equation}
    \log_{10}(q) = b(z)\log_{10}(\mu_*) + (\delta_2 - \delta_1),
\end{equation}
where $\delta_i$ are independent scatter terms with zero mean. Since observed slopes $b(z)$ are of order unity \cite{Pacucci2024, Kormendy2013} and $E[\delta_2 - \delta_1] = 0$, the approximation $q \approx \mu_*$ captures the dominant mean behavior, neglecting the intrinsic scatter $\sigma \approx 0.3-0.7$ dex. This approximation defines the validation target used in this paper; a complete model of SMBH binary pairing physics would require additional assumptions beyond this validation setup.

\subsection{Observational Model Parameters}
We use the LISA noise PSD from \cite{Robson2018} with arm length $L = 2.5 \times 10^9$ m, high-frequency OMS noise $(1.5 \times 10^{-11} \, \text{m})^2 \text{Hz}^{-1}$, and acceleration noise $(3 \times 10^{-15} \, \text{ms}^{-2})^2 \text{Hz}^{-1}$.

\section{Bounded Marginal Distributions}
\label{appendix:bounded_marginals}

This appendix records the bounded marginal densities used by the compound-mixture population model. For a physical coordinate $x\in[a,b]$, define $y=(x-a)/(b-a)$. A Kumaraswamy marginal with shape parameters $\alpha,\beta>0$ has
\begin{equation}
\begin{aligned}
f(x)&=\frac{\alpha\beta}{b-a}y^{\alpha-1}\left(1-y^\alpha\right)^{\beta-1},\\
F(x)&=1-\left(1-y^\alpha\right)^\beta .
\end{aligned}
\end{equation}
A normal density truncated to $[a,b]$ has
\begin{equation}
f(x)=
\frac{\phi[(x-\mu)/\sigma]}
{\sigma\{\Phi[(b-\mu)/\sigma]-\Phi[(a-\mu)/\sigma]\}},
\end{equation}
with the corresponding CDF obtained by replacing the numerator with
$\Phi[(x-\mu)/\sigma]-\Phi[(a-\mu)/\sigma]$. For a Pareto-like marginal on $[a,b]$ with power-law index $\gamma$, we use the normalized form
\begin{equation}
f(x)=
\begin{cases}
\dfrac{(1-\gamma)x^{-\gamma}}{b^{1-\gamma}-a^{1-\gamma}}, & \gamma\neq1,\\[0.8em]
\dfrac{1}{x\log(b/a)}, & \gamma=1,
\end{cases}
\end{equation}
with the analogous integrated CDF. These marginals are evaluated only on the declared support of each application.

\section{LISA Multistart Selected-Medoid Summary}
\label{appendix:lisa_medoid_summary}

Table~\ref{tab:lisa_selected_medoids} gives the selected medoid summary for each mock catalog in the repeated-validation study. The entries document the truth-blind selection procedure. The stable-cluster size is determined before consulting the true counts or shape distances. The coverage columns and expected intrinsic count are validation quantities evaluated after selection.

\begin{table}[t]
\centering
\caption{Selected medoids in the LISA repeated-validation study. Coverage columns indicate whether the selected medoid's 90\% interval contains the true intrinsic count $N_{\text{int}}$ and realized observed count $N_{\text{obs}}$.}
\label{tab:lisa_selected_medoids}
\scriptsize
\setlength{\tabcolsep}{3pt}
\begin{ruledtabular}
\begin{tabular}{lcccc}
Cat. & $n_c$ & $N_{\text{int}}$ & $N_{\text{obs}}$ & $N_{\rm int}^{\rm exp}$ \\ \hline
N20-r0 & 5 & yes & yes & 20.129 \\
N20-r1 & 5 & yes & yes & 18.440 \\
N20-r2 & 5 & yes & yes & 19.870 \\
N20-r3 & 4 & yes & yes & 19.388 \\
N20-r4 & 5 & yes & yes & 22.421 \\
N50-r0 & 5 & yes & yes & 45.228 \\
N50-r1 & 5 & yes & yes & 47.452 \\
N50-r2 & 5 & yes & yes & 43.179 \\
N50-r3 & 5 & yes & yes & 50.445 \\
N50-r4 & 5 & yes & yes & 45.713 \\
N200-r0 & 5 & yes & yes & 198.696 \\
N200-r1 & 5 & yes & yes & 213.810 \\
N200-r2 & 4 & yes & yes & 188.224 \\
N200-r3 & 5 & yes & yes & 191.980 \\
N200-r4 & 5 & yes & yes & 191.200 \\
N500-r0 & 5 & yes & yes & 513.947 \\
N500-r1 & 5 & yes & yes & 509.345 \\
N500-r2 & 5 & yes & yes & 483.605 \\
N500-r3 & 5 & yes & yes & 469.182 \\
N500-r4 & 5 & yes & yes & 501.085 \\
\end{tabular}
\end{ruledtabular}
\end{table}

\section{Parametric Comparison in Controlled Tests}
\label{appendix:parametric_comparison}

This appendix records the limited parametric comparison summarized in Sec.~\ref{sec:comparison}. The comparison is designed as a focused test of model flexibility in a LISA-like setting. The test is deliberately simplified to two dimensions by fixing $q=1$, and both the flexible model and the parametric baseline are applied to mock catalogs with LISA selection effects. The point of the exercise is to separate two regimes: one in which the parametric model is correctly specified, and one in which the injected population has structure outside the assumed parametric family.

In the correctly specified case, the parametric baseline is given the true functional form of the merger-rate density and infers its governing hyperparameters. As expected, this model achieves a near-exact reconstruction. The flexible model, without being supplied the functional form, still recovers the broad morphology with high correlation against the truth. The numerical comparison is shown in Table~\ref{tab:comparison_correct}. This result is useful because it confirms that the flexible density can perform reasonably even when the simple model is correct, while the simple model remains the most efficient description in that regime.

\begin{table}[t]
\centering
\caption{Diagnostic comparison when the parametric model is correctly specified. The baseline is given the true functional form, while the flexible model is fit without that functional information.}
\label{tab:comparison_correct}
\begin{ruledtabular}
\begin{tabular}{lcc}
Comparison & JS divergence & Pearson correlation \\ \hline
Flexible vs truth & 0.0926 & 0.9395 \\
Parametric vs truth & 0.0293 & 0.9948 \\
Flexible vs parametric & 0.0905 & 0.9433 \\
\end{tabular}
\end{ruledtabular}
\end{table}

In the misspecified case, the mock catalog is generated from an alternative ``bimodal mass plus cosmic-noon peak'' population. The injected distribution contains two mass features, centered at $\log_{10}(M_\bullet/M_\odot)=5.5$ and 7.5, and a redshift evolution that rises and then declines around $z\simeq2.5$. These features are intentionally difficult for the restricted parametric form. Table~\ref{tab:comparison_misspecified} shows that the flexible density follows the injected morphology more closely in this diagnostic, reducing the JS divergence from 0.2211 to 0.1095 while maintaining a Pearson correlation of 0.9537. This supports the use of the proposed density model as a complement to parametric analyses when the appropriate functional form is uncertain.

\begin{table}[t]
\centering
\caption{Diagnostic comparison when the parametric model is misspecified. The injected population contains a bimodal mass distribution and nonmonotonic redshift evolution.}
\label{tab:comparison_misspecified}
\begin{ruledtabular}
\begin{tabular}{lcc}
Comparison & JS divergence & Pearson correlation \\ \hline
Flexible vs truth & 0.1095 & 0.9537 \\
Parametric vs truth & 0.2211 & 0.8397 \\
Flexible vs parametric & 0.2331 & 0.8442 \\
\end{tabular}
\end{ruledtabular}
\end{table}

The comparison has two important limitations. It omits the full three-dimensional LISA validation used in the main text and comparisons against optimized implementations of all available sampling methods. Its role is narrower: it demonstrates that the proposed population density can behave sensibly when a simple parametric model is correct, and can retain additional morphology when the injected population lies outside the assumed parametric family.

\section{Analysis Configurations and Stability Criteria}
\label{appendix:configuration_qc}

This appendix summarizes the analysis configurations used in the two applications. In both cases the likelihood uses the same event-level posterior-prior reweighting and a separate population-level selection correction. The differences lie in the event-level uncertainty model, the selection-function representation, the population-density basis, and the stability criteria used to decide whether a fitted morphology is sufficiently reproducible across independent starts.

For the LISA validation, the data are controlled mock catalogs of supermassive black hole binaries with known intrinsic truth. Event-level uncertainty is represented by Fisher-approximation posterior samples, and the detection probability is evaluated with the same differentiable SNR surrogate used to generate the mock detections. The population density is a streamlined model with two Kumaraswamy mixture components. Each catalog is fitted from five independent training starts. The stability rule is based on truth-blind summaries of the trained starts; selected medoids are then validated against the known counts and shape distances. Because the truth is known and the survey construction is controlled, this application can test both shape recovery and expected-count normalization. The repeated-validation evidence reported in the main text is therefore a validation of the full controlled inference procedure, including posterior reweighting, selection correction, multistart stability selection, and post-selection coverage.

For the LVK GWTC-3 application, the data are public posterior samples for confident O1/O2/O3-era compact-binary events after the BBH support cuts described in Sec.~\ref{sec:lvk_data_selection}. Event-level uncertainty comes from the GWOSC parameter-estimation products, and each event is reweighted using the provided prior information. The selection correction is computed from the official O1/O2/O3 injection campaign with the O3 any-IFAR recovery mask and the final support cuts. The population density uses a two-resolution basis in $\log_{10}(m_1/M_\odot)$ together with axis-spline Kumaraswamy and truncated-normal components, so that the lower-mass region can be compared at useful resolution while preserving the broader high-mass support. The LVK stability selection uses truth-blind shape summaries and start-to-start distances.

The separation between stability selection and subsequent comparison is important in both applications. Representative solutions are selected before post-selection checks or external comparisons are performed. This procedure reports results only after independent starts agree on a stable solution, while preserving a clear distinction between numerical stability and scientific interpretation.

\bibliography{references}

\end{document}